\documentclass[sigconf]{acmart}

\copyrightyear{2024}
\acmYear{2024}
\setcopyright{acmlicensed}\acmConference[MSR '24]{21st International Conference on Mining Software Repositories}{April 15--16, 2024}{Lisbon, Portugal}
\acmBooktitle{21st International Conference on Mining Software Repositories (MSR '24), April 15--16, 2024, Lisbon, Portugal}
\acmDOI{10.1145/3643991.3644912}
\acmISBN{979-8-4007-0587-8/24/04}

\usepackage{adjustbox}
\usepackage{listings}
\usepackage{rotating}
\usepackage{tablefootnote}

\acmConference[MSR 2024]{21st International Conference on Mining Software Repositories}{April 2024}{Lisbon, Portugal}
\definecolor{lightgray}{rgb}{.9,.9,.9}
\definecolor{darkgray}{rgb}{.4,.4,.4}
\definecolor{purple}{rgb}{0.65, 0.12, 0.82}

\lstdefinelanguage{JavaScript}{
  keywords={typeof, new, catch, function, return, null, catch, switch, var, if, in, while, do, else, case, break, let, const, continue},
  keywordstyle=\color{blue}\bfseries,
  ndkeywords={class, export, boolean, throw, implements, import, this, false},
  ndkeywordstyle=\color{darkgray}\bfseries,
  identifierstyle=\color{black},
  sensitive=false,
  comment=[l]{//},
  morecomment=[s]{/*}{*/},
  commentstyle=\color{purple}\ttfamily,
  stringstyle=\color{red}\ttfamily,
  morestring=[b]',
  morestring=[b]"
}

\colorlet{punct}{red!60!black}
\definecolor{background}{HTML}{EEEEEE}
\definecolor{delim}{RGB}{20,105,176}
\colorlet{numb}{magenta!60!black}

\lstdefinelanguage{json}{
    basicstyle=\normalfont\ttfamily,
    numbers=left,
    numberstyle=\scriptsize,
    stepnumber=1,
    numbersep=8pt,
    showstringspaces=false,
    breaklines=true,
    frame=lines,
    literate=
     *{0}{{{\color{numb}0}}}{1}
      {1}{{{\color{numb}1}}}{1}
      {2}{{{\color{numb}2}}}{1}
      {3}{{{\color{numb}3}}}{1}
      {4}{{{\color{numb}4}}}{1}
      {5}{{{\color{numb}5}}}{1}
      {6}{{{\color{numb}6}}}{1}
      {7}{{{\color{numb}7}}}{1}
      {8}{{{\color{numb}8}}}{1}
      {9}{{{\color{numb}9}}}{1}
      {:}{{{\color{punct}{:}}}}{1}
      {,}{{{\color{punct}{,}}}}{1}
      {\{}{{{\color{delim}{\{}}}}{1}
      {\}}{{{\color{delim}{\}}}}}{1}
      {[}{{{\color{delim}{[}}}}{1}
      {]}{{{\color{delim}{]}}}}{1},
}

\begin{document}

\title{CrashJS: A NodeJS Benchmark for Automated Crash Reproduction}

\author{Philip Oliver}
\email{philip.oliver@vuw.ac.nz}
\orcid{0000-0003-2989-8478}
\affiliation{%
  \institution{Victoria University of Wellington}
  \city{Wellington}
  \country{New Zealand}
}

\author{Jens Dietrich}
\email{jens.dietrich@vuw.ac.nz}
\thanks{The work of the second author was supported by a gift from Oracle Labs Australia.}
\affiliation{%
  \institution{Victoria University of Wellington}
  \city{Wellington}
  \country{New Zealand}
}

\author{Craig Anslow}
\email{craig.anslow@ecs.vuw.ac.nz}
\affiliation{%
  \institution{Victoria University of Wellington}
  \city{Wellington}
  \country{New Zealand}
}

\author{Michael Homer}
\email{michael.homer@vuw.ac.nz}
\affiliation{%
  \institution{Victoria University of Wellington}
  \city{Wellington}
  \country{New Zealand}
}
\renewcommand{\shortauthors}{Oliver, et al.}

\begin{abstract}
Software bugs often lead to software crashes, which cost US companies upwards of \$2.08 trillion annually.
Automated Crash Reproduction (ACR) aims to generate unit tests that successfully reproduce a crash.
The goal of ACR is to aid developers with debugging, providing them with another tool to locate where a bug is in a program. 
The main approach ACR currently takes is to replicate a stack trace from an error thrown within a program.
Currently, ACR has been developed for C, Java, and Python, but there are no tools targeting JavaScript programs.
To aid the development of JavaScript ACR tools, we propose CrashJS: a benchmark dataset of 453 Node.js crashes from several sources.
CrashJS includes a mix of real-world and synthesised tests, multiple projects, and different levels of complexity for both crashes and target programs.
\end{abstract}


\keywords{Automated Crash Reproduction, Benchmark, Data Collection, Dataset, Software Testing, Test Generation}

\maketitle

\section{Introduction}

`Program testing can be used to show the presence of bugs, but never to show their absence.'
Dijkstra's famous quote implies that while testing can show that a program is buggy, it cannot show that no bugs exist~\cite{dijkstra}.
Bugs show themselves in several manners, with one of the more destructive being software crashes.
Bugs that lead to software crashes can occur in many ways, from incorrect code to changes to dependencies, and security vulnerabilities.
Crashes are a significant cost to companies; the Consortium for Information and Software Quality reported in 2021 that crashes cost US companies more than \$2.08 trillion annually~\cite{cisq}.
Crashes typically result in some form of post-crash data being saved to allow for debugging, this often takes the form of a stack trace.
While developers can use this information to debug the crash, it can often take a significant amount of the developers' time to isolate and fix the cause of the crash.

Automated Crash Reproduction (ACR) is a relatively new idea within software engineering, which aims to reproduce the cause of a crash, allowing developers another point of reference to understand why a crash is occurring.
The reproductions from ACR tools (a tool that aims to reproduce a crash based on limited input) usually take the form of a unit test, which can be included within the test suite for the crashing software as a regression test to ensure the bug is not reintroduced in later versions of the software.
In the last few years, several tools have been created that focus on software crash reproduction.
The majority of these tools target statically typed languages such as Java or C for their crash reproductions~\cite{soltani2016,botsing2020,nayrolles2015,chen2015}, with a notable exception: Beacon, which targets the dynamically typed language Python~\cite{beacon2021}.
However, to the best of our knowledge, there currently are no ACR tools targeting JavaScript.
An issue we have identified in the ACR space is the lack of consistent benchmark datasets to allow direct comparison between existing and forthcoming ACR tools.
As identified by Tempero~\emph{et al.}, datasets that allow reproducible studies and ease of comparison between similar analyses are crucial for the future of empirical software engineering~\cite{qualitasCorpus}.

In this paper, we present CrashJS: a first-of-its-kind benchmark of 453 Node.js crashes collected from multiple sources, some of which have been collected using a novel technique of extracting crashes from test generation tools.
We believe the existence of CrashJS will be crucial for the development and evaluation of new JavaScript tools for ACR and will allow for easier comparisons and development of research within this area.
Furthermore, due to the lack of high-quality GitHub crashes for JavaScript, we present several novel techniques for utilising existing tools for JavaScript to extract crashes (BugsJS, SecBench.js, Syntest-JavaScript).
The contributions this paper presents are: the CrashJS benchmark, artifact, analysis, and a novel approach to extracting crashes from test generation tools.

\section{Existing Benchmarks}

Benchmarks and datasets exist in many forms and for many purposes within software engineering with new datasets published continually.
At MSR 2023 alone, 22 papers were accepted in the Data and Tool Showcase Track across many areas.
Some of these datasets focus on testing various aspects of software~\cite{DeepScenario, microSecEnD, Wasmizer, HPCBugs, HasBugs, SnapshotTesting, ABLoTS}, others focus on aspects of the software development lifecycle~\cite{HelmDataset, GovernanceDataset, Defectors, PyMigBench, GitHubActivities, DACOS, PENTACET, DocMine, SecretBench, GIRTData}, and others look at aspects of Artificial Intelligence and Machine Learning~\cite{PTMTorrent, LLMSecEval, LLMGenderBias}.
Aside from these new datasets, several older, more established datasets include the Qualitas Corpus~\cite{qualitasCorpus}, Da Capo~\cite{DaCapo}, the Software-artifact Infrastructure Repository~\cite{SIR}, and Defects4J~\cite{defects4j}, among others.

Currently, there are few benchmarks for ACR tools.
A particular issue in this area is that separate benchmarks must be created for different programming languages.
An ACR tool targeting Java cannot use a benchmark for JavaScript, for example.
In this section, we discuss several benchmarks for ACR and some JavaScript benchmarks for other purposes such as test generation and security vulnerability analysis.

\subsection{Java Benchmarks}

We identified three main benchmarks used for Java ACR tools.
Only one of these benchmarks (JCrashPack) was created as a benchmark to be used for comparisons by different tools.

STAR is an ACR tool proposed by Chen and Kim to tackle two main issues with other crash-reproduction techniques: path explosion and object creation~\cite{chen2015}.
STAR was evaluated on a set of 52 crashes from Apache Commons Collections (ACC), Apache ANT, and Apache Log4J.
12 of these crashes are from ACC, 20 from ANT, and 20 from Log4J.
STAR was able to exploit 31 of the 52 crashes, with 22 of these being considered useful for revealing the bugs.
Challenges discovered for reproducing crashes were identified as reliance on environmental dependencies (36.7\% of unreproducible crashes), SMT solver limitations (23.3\%), concurrency and non-determinism (16.7\%), path explosion (6.7\%), and other challenges such as reflection (10\%).
These challenges are all areas for further work.
Furthermore, STAR is restricted by the type of exception thrown, with no support for crashes such as \mbox{\emph{ClassNotFoundException}} or \emph{InterruptedException}.

Several tools have been evaluated using the STAR benchmark.
In 2017, Soltani \emph{et al.} presented EvoCrash, an evolutionary tool that leverages a stack trace to reduce the search space~\cite{soltaniGuidedGenetic}.
EvoCrash reproduces crashes for the Java programming language.
To reproduce crashes, EvoCrash uses a guided genetic algorithm with a custom fitness function developed by Soltani \emph{et al.} and uses EvoSuite as an engine to run an evolutionary search and to generate tests~\cite{soltani2016, soltaniGuidedGenetic, evosuite2010}.
EvoCrash replicated 41 of 50 (82\%) crashes from Apache Commons Collections (ACC), Apache Ant (ANT), and Apache Log4j~\cite{soltaniGuidedGenetic} from the STAR dataset.
MuCrash, another Java ACR tool uses only the ACC crashes from the STAR benchmark~\cite{mucrash}.

Nayrolles~\emph{et al.} selected 20 crashes to evaluate JCHARMING.
These crashes were selected from Apache ANT, ArgoUML, Dnsjava, JfreeChart, Apache Log4J, Mission Control Technologies, and PDFBox.
The authors state that the crashes were randomly selected to avoid the introduction of bias but include no further information as to where or how the crashes were selected, beyond ensuring that the exception matches a regular expression the authors constructed to ensure only stack traces were collected.

In 2020, Soltani \emph{et al.} introduced JCrashPack, a set of 200 Java crashes, and evaluated EvoCrash on these crashes~\cite{jCrashPack2020}.
JCrashPack is comprised of crashes from several projects: \emph{Apache commons-lang, Apache commons-math, Closure compiler, ElasticSearch, Joda-Time, Mockito}, and \emph{XWiki}.
EvoCrash successfully reproduced 87 of 200 (43.5\%) from the benchmark JCrashPack. 

Another tool, Botsing, uses JCrashPack for its evaluation.
Botsing, like EvoCrash, is a search-based crash reproduction tool for the Java programming language built on top of EvoSuite~\cite{botsing2020}.
Derakhshanfar \emph{et al.} identified the work of Soltani \emph{et al.}~\cite{soltani2020} as showing the ability of evolutionary search for crash reproduction.
Botsing uses an evolutionary search to produce a test case that replicates the crash behaviour.
In addition to the random nature of mutation and crossover inspired by EvoCrash, Botsing also implements seeding mechanisms.
These mechanisms seed object and method calls based on existing tests and models of classes using \emph{test seeding} and \emph{behavioural model seeding}.
Botsing was evaluated on JCrashPack~\cite{jCrashPack2020} and achieved successful reproduction of 66 of the 124 (53.2\%) crashes without the use of the seeding strategies and 70 (56.5\%) with the model and test seeding mechanisms.
In Derakhshanfar's PhD thesis, they discuss that all the ElasticSearch crashes had been excluded as Botsing was unable to perform the dynamic analysis while executing the ElasticSearch crashes~\cite{pouria2021}.
It is stated that this occurred due to ``the technical difficulty of running ElasticSearch tests by the EvoSuite test executor.''
Including the ElasticSearch tests, Botsing is capable of reproducing 33\% of JCrashPack.
As both EvoCrash and Botsing were evaluated on JCrashPack, a direct comparison of these tools can be made, with EvoCrash reproducing 87 out of 200 crashes, while Botsing could reproduce 70 of the 200 crashes.

\subsection{Python Benchmarks}

There is currently only one tool for ACR which targets the Python language.
Beacon was proposed as an adaptation of the approach EvoCrash uses for ACR~\cite{beacon2021}.
While Beacon does not use a specific benchmark it is, to the best of our knowledge,  the only tool currently performing ACR for a dynamically typed language.
Bergel and Mu\~{n}oz used only three crashes to assess Beacon~\cite{beacon2021}.
The first crash uses only the Python standard library, the second \emph{NumPy}, and the third \emph{PyYAML}.
The first crash is described as an \emph{erroneous list extension}.
This crash has a function that should return a value; however, the crash returns the Python type $None$, which leads to an $AttributeError$.
The second crash is thrown when a buffer is incorrectly used through a buffer size that is too small, or a buffer size that is not a multiple of 16.
The final crash is a $ScannerError$ which is thrown when a string in a YAML document starts with an ampersand ($\&$) and is not a correct reference.
Beacon also requires the message from this crash to contain ``while scanning an anchor,'' making this crash more specific than just looking for the $ScannerError$.
The evaluation was performed 32 times with a population size of 64.
When evaluated, Beacon could effectively reproduce the first two crashes, with 93.75\% and 100\% accuracy in 105.31ms and 149.04ms, respectively.
The \emph{PyYAML} crash was significantly harder for Beacon, with 71.88\% accuracy in 9043.39ms.

\subsection{JavaScript Benchmarks}

There are currently no benchmarks for ACR in JavaScript.
The following two benchmarks are for bug analysis and test generation tools~\cite{BugsJS, syntest2022}.
The third benchmark is a collection of real-world vulnerabilities within JavaScript packages~\cite{secbenchjs}.
All of these benchmarks are unsuitable in their current state for ACR as they do not contain crashes or stack traces produced by crashes.
These three tools are leveraged in the creation of the CrashJS benchmark.

BugsJS was presented by Gyimesi~\emph{et al.} in 2019 as a benchmark of 453 JavaScript bugs, fixes, and test suite updates~\cite{BugsJS}.
These bugs were collected from 10 Node.js projects that use the Mocha testing framework.
The 10 projects used in BugsJS are Bower, Eslint, Express, Hessian.js, Hexo, Karma, Mongoose, Node-Redis, Pencilblue, and Shields.
Gyimesi~\emph{et al.} selected these projects as they had the highest star count on GitHub, were server-side Node.js projects with a large number of commits and were actively maintained.
Projects were selected only when bug reports were included in issue tracking on GitHub.

For each project, Gyimesi~\emph{et al.} selected bugs from the list of closed issues where there is a link between the issue and a bug-fixing commit.
For these issues, only those that had test changes present were included.
The resulting issues were then assessed against criteria set out by Gyimesi~\emph{et al}: isolation, complexity, dependency, relevant changes, and refactoring.
These criteria ensure that patches in BugsJS are high quality but not so complex as to require much domain-specific knowledge to fix.
From an original list of 795 commits, the authors used these criteria to reduce the list to 542.
Finally, the authors used dynamic validation to ensure that at least one of the tests added in the patched version fail when executed on the buggy version of the code.
After this analysis, the list of bugs was further reduced to the 453 included in BugsJS.

Syntest-JavaScript is an automated test generation tool for \\ JavaScript and TypeScript programs~\cite{syntest2022}.
Similar to tools such as EvoSuite~\cite{evosuite2010}, Syntest-JavaScript aims to maximise code coverage as its main goal.
Stallenberg~\emph{et al.} constructed a dataset of 98 units under test for Syntest-JavaScript.
Five JavaScript projects with high star counts were selected from GitHub: Commander.js, Express, Moment.js, Javascript Algorithms, and Lodash.
The criteria for inclusion in the Syntest-JavaScript benchmark were that the unit must be exported and the unit must have a Cyclomatic Complexity greater or equal to 2, as calculated by Plato\footnote{https://github.com/es-analysis/plato}.
The authors also noted that two files in Commander.js would terminate the process and had to be excluded from the benchmark as they would exit the testing tool.

Bhuiyan~\emph{et al.} presented SecBench.js in 2023 as a server-side JavaScript dataset for security vulnerabilities~\cite{secbenchjs}.
The authors describe the need for a successful benchmark to be realistic, executable, two-sided, and vetted (or confirmed).
A two-sided vulnerability means the item in the benchmark contains the vulnerable and fixed versions of the vulnerability.
This requirement allows confirmation of detection and mitigation tools and further insight into how vulnerabilities are fixed.
Bhuiyan~\emph{et al.} collected 600 vulnerabilities from Snyk, GitHub Advisories, and Huntr.dev.
The vulnerabilities collected fall into 5 categories: code injection (40), command injection (101), path traversal (169), prototype pollution (192), and ReDoS (98).
Code injection allows arbitrary code execution while command injection allows arbitrary CLI commands to be executed.
Path traversal allows an arbitrary path in the file system to be read.
Prototype pollution allows attributes to be added or modified on prototypes such as the built-in JavaScript Object class.
Finally, ReDoS vulnerabilities use regular expressions which take a long time to execute, thus overloading system resources and causing a denial of service.
The 600 vulnerabilities in SecBench.js are represented as executable Jest test cases.
These test cases are executed on the target units and show the presence of the vulnerabilities.

\begin{table}[htb]
\centering
\caption{Existing Benchmarks}
\label{T:existingBenchmarks}
\begin{tabular}{ | c | c | c | c | } 
 \hline
 Benchmark & Language & Purpose & Size \\
 \hline
 \hline
 STAR & Java & ACR & 52 \\
 \hline
 JCHARMING & Java & ACR & 20 \\ 
 \hline 
 JCrashPack & Java & ACR & 200 \\
 \hline 
 Defects4J & Java & Bug Detection & 750 \\
 \hline
 Beacon & Python & ACR & 3 \\
 \hline 
 BugsJS & JS & Bug Detection & 453 \\
 \hline 
 Syntest-JS & JS & Test Generation & 98 \\
 \hline 
 SecBench.js & JS & Security Vulnerabilities & 600 \\
 \hline
 \hline
 CrashJS & JS & ACR & 453 \\
 \hline
\end{tabular}
\end{table}

\section{CrashJS}

We propose CrashJS: a benchmark of 453 Node.js crashes specifically collected for use by tools for ACR, consisting of crashes from four primary locations: GitHub\footnote{https://github.com}, BugsJS~\cite{BugsJS}, Syntest\-JavaScript~\cite{syntest2022}, and SecBench.js~\cite{secbenchjs}.
71 crashes are collected from GitHub, 90 from BugsJS, 275 from Syntest-JavaScript, and 17 from SecBench.js.
The CrashJS artifact including crashes, collection scripts, and analysis scripts can be found at \url{https://zenodo.org/doi/10.5281/zenodo.10530514}~\cite{crashjsDataset}.

\subsection{Collection Method} \label{SS:collectionMethod}

To create a suitably sized benchmark of crashes, we have used four main approaches.
The first of these approaches is to collect crashes from issues on popular GitHub projects.
The second is to use BugsJS, a benchmark of JavaScript bugs collected by Gyimesi~\emph{et al.}~\cite{BugsJS}.
Thirdly, we leveraged reported JavaScript security vulnerabilities from Secbench.js~\cite{secbenchjs}.
Finally, we collected stack traces from the tool Syntest-Javascript presented by Stallenberg~\emph{et al.}~\cite{syntest2022}.
For SecBench.js and Syntest-JavaScript crashes we have included the test case which generated the error.
This will allow users of CrashJS to compare generated test cases to the initial cause of the error, thus providing an oracle and an opportunity to assess the usefulness of generated test cases.

There are several steps involved in collecting each crash from GitHub.
The initial step is selecting projects on GitHub from which to collect crashes.
Projects under the \emph{JavaScript} topic with the main language identified as JavaScript were explored.
The resulting projects were sorted by the number of \emph{stars}, GitHub's in-built measure of popularity.
This GitHub search can be reproduced using this URL: \url{https://github.com/topics/javascript?l=javascript&o=desc&s=stars}.
Projects from this search that are client-side JavaScript frameworks or tools, such as \emph{React, Vue}, and \emph{Bootstrap} were excluded as these projects are built to run on web browsers, rather than Node.js.
Once projects were selected, we searched through the issue trackers for these projects.
When selecting issues, we used closed issues only and sorted these by most recent.
This allowed us to confirm if these issues were relevant to the project, as decided by the moderators' comments.
Using recent issues ensured that finding the correct version of the project (if it was not provided), and required dependencies was significantly easier.
In some cases, manual verification of project and dependency versions was required.
This involved cross-referencing functions and line numbers present in the crash stack trace with the code in multiple versions of dependencies to find the versions which matched the stack traces.
As we collected crashes for ACR tools, most of which require a stack trace, we collected only issues including a stack trace.
Table~\ref{T:githubCrashes} provides a summary of the total number of crashes collected for each GitHub project.

\begin{table}[htb]
\centering
\caption{Crashes Collected from GitHub}
\vspace{-1em}
\label{T:githubCrashes}
\begin{tabular}{ | c | c | } 
 \hline
 GitHub Project & No. Crashes Collected \\
 \hline
 \hline
 atom\tablefootnote{\label{N:atom}\url{https://github.com/atom/atom}} & 17 \\
 \hline
 eslint\tablefootnote{\label{N:eslint}\url{https://github.com/eslint/eslint}} & 9 \\
 \hline
 express\tablefootnote{\label{N:express}\url{https://github.com/expressjs/express}} & 13 \\ 
 \hline
 http-server\tablefootnote{\label{N:http-server}\url{https://github.com/http-party/http-server}} & 9 \\
 \hline
 node\tablefootnote{\label{N:node}\url{https://github.com/nodejs/node}} & 11 \\ 
 \hline
 standard\tablefootnote{\label{N:standard}\url{https://github.com/standard/standard}} & 3 \\
 \hline
 webpack\tablefootnote{\label{N:webpack}\url{https://github.com/webpack/webpack}} & 9 \\ 
 \hline
\end{tabular}
\end{table}

The BugsJS benchmark contains several scripts to checkout dependencies from GitHub and run tests for the bugs included in the benchmark.
The tests are run using the Istanbul test runner and results are collected and stored in a JSON file.
The process we used for discovering and collecting crashes from BugsJS follows:
Running the tests in BugsJS required the use of an older version of Node to ensure these ran correctly; we used Node v14.17.4.
First, the tests for the buggy versions of each bug were run to create JSON files containing information about the test executions.
These resulting JSON files include test failures and errors that were encountered while running the tests.
We also saved the \texttt{package.json} file for each test so we could extract dependency versions and other information about each project for analysis and to create the crash files discussed in Section~\ref{SS:collectionMethod}.
These generated JSON files contained the errors encountered when running the BugsJS tests.
We wrote a script to extract these stack traces from the generated JSON file to run deduplication on these crashes and to reformat them into the format from Listing~\ref{L:crashObject} for CrashJS.
We also copied code directories which might be required in reproducing the crash.
For example, some of the crashes for Express occur in files in a folder called \texttt{test}. 
This folder is not included in a production build of Express, so we decided to include these files so the main project and dependencies can be downloaded by ACR tools using the NPM package manager.
Finally, the created log, crash, and test files were added to the CrashJS dataset.
The scripts for this collection can be found in the CrashJS artefact.

Aside from manually collecting crashes from GitHub, SecBench.js was the most time-consuming and difficult origin for collecting crashes.
This was mostly in part due to each vulnerability within SecBench.js being individually represented by a Jest test case.
These test cases required modification to produce an error which could be used as a target for ACR.
Of the 5 vulnerability classes represented in SecBench.js (code injection, command injection, path traversal, prototype pollution, and ReDoS), we were successful in creating crashes in code injection vulnerabilities.

\begin{table}[htb]
\centering
\caption{Crashes Collected from SecBench.js}
\label{T:secbench}
\begin{tabular}{ | c | c | c | } 
 \hline
 SecBench.js Project & No. Crashes Collected & Version(s) \\
 \hline
 \hline
 access-policy & 1 & 3.1.0 \\
 \hline
 json-ptr & 1 & 2.0.0 \\
 \hline
 kmc & 1 & 1.2.2 \\
 \hline
 m-log & 1 & 0.0.1 \\
 \hline
 mathjs & 2 & 3.9.0, 3.10.3 \\
 \hline
 modjs & 1 & 0.4.0 \\
 \hline
 modulify & 1 & 0.1.0 \\
 \hline
 mol-proto & 1 & 0.1.3 \\
 \hline
 mongoosemask & 1 & 0.0.6 \\
 \hline
 node-extends & 1 & 0.2.0 \\
 \hline
 node-rules & 1 & 3.0.0 \\
 \hline
 node-serialize & 1 & 0.0.3 \\
 \hline
 realms-shim & 1 & 1.1.0 \\
 \hline
 serialize-to-js & 1 & 0.5.0 \\
 \hline
 thenify & 1 & 3.3.0 \\
 \hline
 underscore & 1 & 1.13.0 \\
 \hline
\end{tabular}
\end{table}

Command injection vulnerabilities directly call CLI commands external from the executing JavaScript program. 
As this attack is occurring externally, we do not consider it possible to create a crash from these vulnerabilities. 
An approach we attempted was to kill the JavaScript test from the CLI; however, this approach does not produce a stack trace and thus has been excluded from CrashJS.

Path traversal vulnerabilities don't call code, as they are focused on accessing files on a file system external to the JavaScript execution. 
Due to this, we cannot extract stack traces and therefore crashes from these vulnerabilities, so path traversal vulnerabilities have been excluded from CrashJS.

Prototype pollution vulnerabilities can lead to arbitrary code execution and could be used for extracting crashes.
However, crashes for these vulnerabilities require an error to be inserted into the modified prototype function and then called separately.
The stack traces which occur only contain stack frames which are testing framework setup functions and functions from the test itself.
No frames from the dependency which contains the vulnerability are present in the stack trace.
It was concluded that these stack traces would not be useful in CrashJS as they do not aim for a dependency to be tested by an ACR system and so prototype pollution vulnerabilities have been excluded from CrashJS.

ReDoS vulnerabilities do not create stack traces as they aim to create long-executing processes.
As these vulnerabilities simply overload system resources and do not produce errors, they have been excluded from CrashJS.

Only code injection vulnerabilities are included in CrashJS.
To extract stack traces from these tests, we modified the line of each test which writes a file to the filesystem to instead throw an error.
The code injection tests were then run and the errors were collected, along with the dependencies and versions to be included in CrashJS.
Using this method, we collected 17 code injection crashes from SecBench.js to include in CrashJS.

To the best of our knowledge, the method used here for collecting crashes from Syntest-JavaScript is novel.
We have manipulated the coverage-driven test generation tool to output stack traces of errors it encounters in the test generation process.

The first of the steps to collect crashes from Syntest-JavaScript was to select the benchmark on which to run the tool.
We selected the Syntest-JavaScript-Benchmark project from the Syntest-Framework project on GitHub as of 10 October 2023.
This benchmark consists of projects specifically chosen for evaluating Syntest-JavaScript by generating tests using code coverage as a metric.
To collect crashes from the Syntest-JavaScript benchmark, we modified Syntest-JavaScript to collect and save stack traces that occurred while the tool was improving coverage during its test case generation.
This simply involved printing out any errors that were encountered by the test runner with some surrounding lines to allow for easy identification of where these errors were in the output file.
We also printed the test code which led to the error so that the test file could be included in CrashJS.
Another modification we made was to add \texttt{retainLines:~true} to the Babel configuration for Syntest.
This change means that when the system under test is instrumented the instrumenter will output code on the same line as the original file, ensuring that the line numbers in the stack traces will match between the original and instrumented versions of the code.
We modified several variables within the \texttt{.syntest.json} configuration file to reduce the time taken to extract crashes.
The variables we specified were:
\texttt{search-algorithm: MOSAFamily}, 
\texttt{objective-manager: simple},
\texttt{crossover: javascript-tree},
\texttt{procreation: default},
\texttt{sampler: javascript-random},\\
\texttt{total-time: 600},
\texttt{search-time: 600}, and
\texttt{iterations: 20}.

Crashes randomly generated by Syntest-JavaScript are useful for JS automated crash reproduction because, while the tests generated by Syntest-JavaScript are synthetic, the crashes they produce are real. 
In the fuzzing area of software engineering research, this approach is typically used to ensure that areas of the codebase that are rarely covered are still explored and tested. 
This is useful for JavaScript, in particular, where cultural factors can result in crashes within system boundaries which may not have a direct impact on execution being disregarded as client error~\cite{javascriptErrors}.
Furthermore, the deep and complex dependency graphs produce many internal subsystem boundaries where unexpected interactions can be undetected~\cite{javascriptDependencies}. 
For example, some crashes in Java benchmarks result from nullable-by-default APIs where methods can accept a null value but do not have any form of checking for null, thus resulting in a crash on well-typed input data. 
In JavaScript, this could be interpreted as a violation of an implicit not-null invariant and thus would be regarded as user error (violated preconditions), rather than a reportable bug. 
This results in a selection bias against these types of bug reports for JavaScript as they are not reported.
However, these bugs are still valid errors and more complex assumed or implicit preconditions are also widespread. 
Syntest-JavaScript produces these sorts of crashes, which should still be considered by an automated crash reproduction tool for use in development environments where a developer may violate these implicit requirements but not be alerted to it by an IDE or other tool. 
These inputs may be valid from a library API but be violations specified in the documentation or assumed within the library, thus resulting in a crash.
While Syntest-JavaScript crashes \textit{may} be trivial for ACR tools to reproduce, more metrics can be used than just the crash being reproduced.
For example, users of CrashJS could identify the number of evolutionary epochs required to reproduce any given crash.
This metric could be used to tune an ACR tool to improve the time taken to find a crash compared to Syntest-JavaScript.
Each collection of crashes within the dataset is separate from the others, so developers of future ACR tools can include or exclude particular collections of crashes within CrashJS at their discretion.


\subsection{Deduplication}

The deduplication performed on the crashes present in BugsJS was critical in reducing the number of crashes within CrashJS to a manageable number.
Our deduplication strategy is as follows:
Firstly, identify if a stack trace is the same as one previously encountered.
This stack trace matching would discard test fixture setup information from the comparison, as this information is irrelevant to the actual crash.
Matching occurred line-by-line to ensure each line of the stack trace occurred in the same code file, on the same line, and at the same character within that line.
If the stack trace matches one already present, the error line (error type and error message) was added to a set to ensure duplicate error lines were removed.
The results of this deduplication were printed to a JSON file as each stack trace with a list of error lines for each project so that manual verification could occur.
We then manually checked each stack trace to remove further duplicates of error lines.
The only error lines we removed manually were error messages where a particular value was present in the error message.
For example, if a string was expected at a particular point of the program but received an integer, the error message might include the value of that integer.
If multiple integers are inputted at that point of the program there could be multiple instances of the same error but with different integer values.
These cases were removed so there was only one instance of these errors.
We selected instances of the duplicated crashes with the simplest values; 
for example, if there was a long integer and a single-digit integer we kept the single-digit integer.
Once the JSON files with the stack trace and error lines had been pruned manually, we then ran a script to transform this information into the crash and log files by combining each stack trace with all the remaining error lines for that stack trace to create multiple crashes.
For example, if there was one stack trace with five error lines, this would result in five crashes.
Table~\ref{T:bugsjsCrashes} shows the results of the crash extraction, the significant reduction in the number of crashes after deduplication, and the final number of crashes collected for each project (excluding Pencilblue) from BugsJS.

\begin{table}[htb]
\centering
\caption{Crashes Collected from BugsJS}
\label{T:bugsjsCrashes}
\begin{tabular}{ | c | c | c | c | c | } 
 \hline
 Proj & Coll & Auto Dedup & Man Dedup & \% Decr \\
 \hline
 \hline
 eslint$^{\ref{N:eslint}}$ & 4,595 & 88 & 52 & 98.9\% \\
 \hline
 express$^{\ref{N:express}}$ & 128 & 15 & 15 & 88.3\% \\ 
 \hline
 hexo\tablefootnote{\url{https://github.com/hexojs/hexo}} & 140 & 20 & 20 & 85.7\% \\
 \hline
 pencilblue\tablefootnote{\url{https://github.com/pencilblue/pencilblue}} & 3 & 3 & 3 & 0\% \\ 
 \hline
 \hline
 Total & 4,866 & 126 & 90 & 98.2\% \\
 \hline
\end{tabular}
\end{table}

When performing deduplication for the BugsJS dataset, we considered calculating the Cyclomatic Complexity Number (CCN) for each duplicated crash and selecting the least complex option.
Initially, it does not seem that crashes with the same stack trace would have different complexities.
However, while the line numbers for each frame within the stack trace remain the same, other files within the target project could have changed, thus changing the complexity of the program.
We decided not to implement this metric due to the processing time to calculate this for every extracted crash, and the overall low impact on the resulting crashes.
As will be seen in the following sections and figures, the CCN for each project varies little and the variation already present in the dataset is sufficient to provide a comprehensive benchmark for JavaScript ACR tools.
This does not apply to the Syntest-JavaScript dataset, as all crashes extracted from each project are from the same version of the project.

We performed the same deduplication strategy on the crashes collected from Syntest-JavaScript as we did for BugsJS.
Due to the evolutionary nature of the approach Syntest-JavaScript uses to generate tests, the total number of tests generated and errors extracted was significantly higher than any of the other approaches used in this paper.
Table~\ref{T:syntestCrashes} shows the results of the crash extraction, the significant reduction in the number of crashes after deduplication, and the final number of crashes collected for each project from Syntest-JavaScript.

\begin{table}[htb]
\centering
\caption{Crashes Collected from Syntest-JavaScript}
\label{T:syntestCrashes}
\begin{tabular}{ | c | c | c | c | c | } 
 \hline
 Project & Coll & Auto Dedup & Man Dedup & \% Dec \\
 \hline
 \hline
 commanderjs\tablefootnote{\url{https://github.com/tj/commander.js}} & 5,521 & 66 & 38 & 99.3\% \\
 \hline
 express$^{\ref{N:express}}$ & 11,992 & 353 & 86 & 99.3\% \\ 
 \hline
 js-algorithms\tablefootnote{\url{https://github.com/trekhleb/javascript-algorithms}} & 32,752 & 145 & 54 & 99.8\% \\
 \hline
 lodash\tablefootnote{\url{https://github.com/lodash/lodash}} & 2,571 & 11 & 11 & 99.6\% \\ 
 \hline
 moment\tablefootnote{\url{https://github.com/moment/moment}} & 26,238 & 130 & 86 & 99.7\% \\
 \hline
 \hline
 Total & 79,074 & 705 & 275 & 99.7\% \\
 \hline
\end{tabular}
\end{table}

\subsection{Benchmark Format}

For all crashes, stack traces are collected and stored in a \emph{log} file, while other information about each crash is stored in a \emph{JSON} file.
The JSON files follow a standard format described by the TypeScript interface model in Listing~\ref{L:crashObject}.
Information about each crash collected includes the issue number, issue title (\emph{info}), URL, the version number of the tool, an optional version for Node.js, and an object of dependency names to versions for other dependencies.
If a crash is collected from GitHub, the issue number is the number of the GitHub issue from where the crash was collected, otherwise, it is a unique identifier for the crash within the source from which the crash was collected.
For GitHub crashes, the URL links to the GitHub issue.

Several options for dependency management are allowed within the crash file.
The first option is the NPM version of the software from where the crash comes.
This option is used in conjunction with the \texttt{requireCrashDependency} option, which informs an ACR tool if it should download the crash target project as a dependency.
For example, an Express crash will require Express as a dependency, so an ACR tool should install the version of Express identified in \texttt{version}.
In contrast, an Atom crash will not require Atom as an NPM dependency (as it is a standalone tool); 
therefore, \texttt{requireCrashDependency} can be set to false and the ACR tool can ignore the project as a dependency.
The \texttt{nodeVersion} option allows specification of a version of Node to use when running the crash;
This ensures that if an ACR tool requires matching between line numbers of Node functions, these will be correct.
The \emph{setup} object offers options to copy files between locations or to download and extract files from a URL.
These options allow for crashes that have dependencies not within an NPM repository (such as whole applications like Atom) to be analysed.

\begin{lstlisting}[caption=Object Definition for Collected Crash, language=json, label=L:crashObject, float=htb]
{
    "issueNumber": number;
    "info": string;
    "url"?: string;
    "version": string;
    "nodeVersion"?: string;
    "dependencies": {
        ["dependencyName": string]: string;
    };
    "seeded"?: boolean;
    "requireCrashDependency"?: boolean;
    "setup"?: {
        "copy"?: {
            "from": string;
            "to": string;
        },
        "download"?: {
            "url": string;
            "unpack": string;
        }
    };
}
\end{lstlisting}

For example, Listing~\ref{L:atomCrash} shows the JSON file for the Atom-22699 crash.
As can be seen, a link to the original GitHub issue is provided along with version numbers.
The setup object is used to download Atom at version 1.57.0 as a tar.gz and the type of compression so the archive file can be extracted.

\begin{lstlisting}[caption=JSON File for Atom-22699 Crash, label=L:atomCrash, float=htb, language=json]
{
  "issueNumber": 22699,
  "info": "Uncaught Error: ENOENT: no such file or directory, stat '$ATOM_HOME/packages\\atom-autocomplete-py...",
  "url": "https://github.com/atom/atom/issues/22699",
  "version": "1.57.0",
  "dependencies": {
    "electron": "9.4.4"
  },
  "setup": {
    "download": {
      "url": "https://github.com/atom/atom/archive/refs/tags/v1.57.0.tar.gz",
      "unpack": "tar.gz"
    }
  }
}

\end{lstlisting}

\section{Analysis}

We have analysed several aspects of the crashes in CrashJS.
Among these analyses, we have looked at the complexity of the crashes collected, the complexity of the target programs, and the distribution of the types of errors within CrashJS.
Crashes from three of our sources (BugsJS, GitHub, Secbench.js) are real-world crashes, while crashes from Syntest-JavaScript are synthetic crashes but generated using a coverage-based testing tool on real-world programs.

\subsection{Crash Complexity}

In collating CrashJS, we aimed to have a variety of crash complexity so that ACR tools can gain insights into how the tools perform on a variety of programs and complex crashes.
Tables~\ref{T:bugsjsStats}, \ref{T:githubStats}, \ref{T:syntestStats}, and  \ref{T:secbenchStats} show the results of analysis of the stack traces for each source of crashes in CrashJS.
The tables present the number of stack traces (\emph{st}), total number of stack frames (\emph{fr}), average number of frames per stack trace ($\overline{\emph{fr}}$), and standard deviation ($\sigma$) for different types of errors and the overall totals.
5 specific errors (excludes \texttt{Error}) are shown: TypeError (TE), AssertionError (AE), RangeError (RE), SyntaxError (SE), and YAMLException (YE).
These errors are only present in the tables if there are any crashes for that type of error.
For example, Table~\ref{T:bugsjsStats} does not include SyntaxError, as there are none in the BugsJS dataset within CrashJS.
The total number of errors by different type are shown in Table~\ref{T:errorTypes}; the most common error is TypeError, with 308 errors, mostly occurring in the Syntest-JavaScript dataset.

The BugsJS dataset contains the most complex crashes, in the Express project, which has an average of 54 frames per crash.
However, the BugsJS dataset also contains the least complex crashes, in the Eslint project, which has an average of 3 frames per crash.
Interestingly, the standard deviation for BugsJS crashes within each project is 0; 
this shows that for each project, every stack trace has the same number of frames: 3 for Eslint, 52 for Express, 16 for Hexo, and 4 for Pencilblue.
The overall average for BugsJS crashes is 14.4 frames per crash, which indicates a reasonably high complexity for the BugsJS dataset within CrashJS.

\begin{table}[htb]
\centering
\caption{BugsJS Crash Statistics}
\label{T:bugsjsStats}
\begin{tabular}{ | c  c | c | c | c | c | c | } 
 \hline
 Project &  & TE & AE & YE & Other & Total \\
 \hline
 \hline
eslint & \emph{st} & 2 & 35 & 3 & 12 & 52 \\
& \emph{fr} & 6 & 105 & 9 & 36 & 156 \\
& $\overline{\emph{fr}}$ & 3.0 & 3.0 & 3.0 & 3.0 & 3.0 \\
& $\sigma$ & 0.0 & 0.0 & 0.0 & 0.0 & 0.0 \\
\hline
express & \emph{st} & 6 & 5 & 0 & 4 & 15 \\
& \emph{fr} & 324 & 270 & 0 & 216 & 810 \\
& $\overline{\emph{fr}}$ & 54.0 & 54.0 & 0 & 54.0 & 54.0 \\
& $\sigma$ & 0.0 & 0.0 & 0.0 & 0.0 & 0.0 \\
\hline
hexo & \emph{st} & 11 & 8 & 0 & 1 & 20 \\
& \emph{fr} & 176 & 128 & 0 & 16 & 320 \\
& $\overline{\emph{fr}}$ & 16.0 & 16.0 & 0 & 16.0 & 16.0 \\
& $\sigma$ & 0.0 & 0.0 & 0.0 & 0.0 & 0.0 \\
\hline
pencilblue & \emph{st} & 0 & 3 & 0 & 0 & 3 \\
& \emph{fr} & 0 & 12 & 0 & 0 & 12 \\
& $\overline{\emph{fr}}$ & 0 & 4.0 & 0 & 0 & 4.0 \\
& $\sigma$ & 0.0 & 0.0 & 0.0 & 0.0 & 0.0 \\
\hline
Total & \emph{st} & 19 & 51 & 3 & 17 & 90 \\
& \emph{fr} & 506 & 515 & 9 & 268 & 1298 \\
& $\overline{\emph{fr}}$ & 26.6 & 10.1 & 3.0 & 15.8 & 14.4 \\
& $\sigma$ & 19.0 & 15.2 & 0.0 & 21.4 & 18.5 \\
\hline
\end{tabular}
\end{table}

Regarding types of errors, it is interesting to note that the Eslint project contains three YAMLExceptions. 
These three crashes are the only crashes within CrashJS with this type of error.
Two of these crashes (9, 31) occur because a configuration file cannot be read.
The other crash (37) occurs due to a duplicate key mapping.
All three of these crashes occur due to incorrect YAML formatting within input data for the tests, which could provide insight into whether an ACR tool can generate YAML.
However, these crashes could hinder an ACR tool. 
If an ACR tool is capable of generating correct YAML, it is possible that these crashes would never be found as the crashes are all parsing errors which would not occur with correct YAML input.
Regardless, these crashes have been left in CrashJS as they could prove useful for researchers looking to incorporate domain-specific languages (DSLs), such as YAML, into input generation for their ACR tools.

The majority of errors in BugsJS are AssertionErrors.
These errors occur due to the nature of BugsJS: each crash comes from running unit tests within a project.
A significant number of the crashes collected from BugsJS are due to changes in expected values because of changes within the codebase or dependencies.
For example, the \texttt{Pencilblue-1} AssertionError occurs where it expects \texttt{application/font-woff} but receives \texttt{font/woff}. 
To a developer, it is obvious that this error arises due to an updated mimetype representation within the codebase somewhere.
Another example of this is in the \texttt{Hexo-4} crash, where a string of HTML is expected but receives a different string of HTML.
Again, to a developer, it is reasonably clear that there has been a change within Hexo to change the HTML generated and the test must be updated.
This crash, like the YAMLExceptions discussed before, could prove an interesting test of whether an ACR tool can generate correct DSL input.

Of the 71 crashes collected from GitHub, the average number of frames for these crashes is 10.2 with a standard deviation of 3.5.
The Atom project contains the most complex crashes, with an average stack length of 12.4, while Express contains the least complex crashes, with an average stack length of 8.8.

\begin{table}[htb]
\centering
\caption{GitHub Crash Statistics}
\label{T:githubStats}
\begin{tabular}{ | c  c | c | c | c | c | } 
 \hline
 Project &  & TE & SE & Other & Total \\
 \hline
 \hline
atom & \emph{st} & 1 & 1 & 15 & 17 \\
& \emph{fr} & 5 & 10 & 196 & 211 \\
& $\overline{\emph{fr}}$ & 5.0 & 10.0 & 13.1 & 12.4 \\
& $\sigma$ & 0.0 & 0.0 & 4.5 & 4.7 \\
\hline
eslint & \emph{st} & 8 & 1 & 0 & 9 \\
& \emph{fr} & 80 & 10 & 0 & 90 \\
& $\overline{\emph{fr}}$ & 10.0 & 10.0 & 0 & 10.0 \\
& $\sigma$ & 0.0 & 0.0 & 0.0 & 0.0 \\
\hline
express & \emph{st} & 8 & 0 & 5 & 13 \\
& \emph{fr} & 73 & 0 & 42 & 115 \\
& $\overline{\emph{fr}}$ & 9.1 & 0 & 8.4 & 8.8 \\
& $\sigma$ & 3.4 & 0.0 & 4.3 & 3.8 \\
\hline
http-server & \emph{st} & 3 & 0 & 6 & 9 \\
& \emph{fr} & 30 & 0 & 56 & 86 \\
& $\overline{\emph{fr}}$ & 10.0 & 0 & 9.3 & 9.6 \\
& $\sigma$ & 0.0 & 0.0 & 1.1 & 1.0 \\
\hline
node & \emph{st} & 5 & 0 & 6 & 11 \\
& \emph{fr} & 48 & 0 & 42 & 90 \\
& $\overline{\emph{fr}}$ & 9.6 & 0 & 7.0 & 8.2 \\
& $\sigma$ & 0.5 & 0.0 & 1.8 & 1.9 \\
\hline
standard & \emph{st} & 2 & 1 & 0 & 3 \\
& \emph{fr} & 20 & 10 & 0 & 30 \\
& $\overline{\emph{fr}}$ & 10.0 & 10.0 & 0 & 10.0 \\
& $\sigma$ & 0.0 & 0.0 & 0.0 & 0.0 \\
\hline
webpack & \emph{st} & 5 & 0 & 4 & 9 \\
& \emph{fr} & 58 & 0 & 44 & 102 \\
& $\overline{\emph{fr}}$ & 11.6 & 0 & 11.0 & 11.3 \\
& $\sigma$ & 3.9 & 0.0 & 1.7 & 3.1 \\
\hline
Total & \emph{st} & 32 & 3 & 36 & 71 \\
& \emph{fr} & 314 & 30 & 380 & 724 \\
& $\overline{\emph{fr}}$ & 9.8 & 10.0 & 10.6 & 10.2 \\
& $\sigma$ & 2.6 & 0.0 & 4.2 & 3.5 \\
\hline

\end{tabular}
\end{table}

The majority of errors in the GitHub dataset are TypeErrors (32), followed by generic Errors (29), with a few other errors such as SyntaxError and URIError.
A significant number of the TypeErrors present in the GitHub dataset are the program under test not being able to access properties on undefined objects.
For example, the \texttt{Eslint-14538} crash occurs in Eslint version 7.25.0 when there is an incomplete variable declaration.
The error given for this crash is: \texttt{TypeError: Cannot read property 'loc' of undefined}, instead of Eslint providing a useful error message such as \texttt{identifier expected}.
Another common TypeError is a missing function.
For example, \texttt{Express-2761} has the error message \texttt{TypeError: dest.end is not a function} for Express version 5.0.0-alpha.2 when attempting to use http2 as a dependency.
This function cannot be found because Express did not support http2 at the time of 5.0.0-alpha.2.
Generic errors include messages such as \texttt{Module did not self-register}, \texttt{no such file or directory}, and \texttt{listen EADDRINUSE 0.0.0.0:8080}.

On first look, it appears that SecBench.js crashes are some of the more complex crashes within CrashJS, with an average stack trace of 12.9 frames.
However, if we dig deeper into the stack traces present in the SecBench.js dataset, we find one crash with a stack length of 100.
If we exclude this crash, we find a mean stack length of 7.19.
The overall median of the SecBench.js dataset stack length is 7 frames.
These statistics show that, excluding the Modjs crash with a length of 100, SecBench.js crashes are not as complex as they first appear.
The Modjs crash is so complex because the majority of the processing for this crash is parsing.
Due to the recursive nature of parsing, the stack trace is significantly larger, as many elements of the program being parsed must be covered.

The majority of errors in the SecBench.js dataset are generic Errors (13), with SyntaxErrors next (3), and a single ReferenceError last.
The majority of these crashes are generic Errors due to how crashes were extracted from SecBench.js.
As these crashes are collected from code injection vulnerabilities within SecBench.js, the errors are collected by injecting a piece of code which throws an Error.
The SyntaxErrors occur in the Node-Rules and Underscore vulnerabilities, where the syntax of throwing an error does not work like in the other crashes collected.
The ReferenceError occurs in the Thenify vulnerability, where \texttt{throw new Error();} is injected as a payload.
However, the error message (thrownewError is not defined) shows the spaces within the payload are replaced and a ReferenceError is thrown instead of an Error.

\begin{table}[htb]
\centering
\caption{SecBench.js Crash Statistics}
\label{T:secbenchStats}
\begin{tabular}{ | c  c | c | c | c | c | } 
 \hline
 Project &  & SE & Error & Other & Total \\
 \hline
 \hline
code-injection & \emph{st} & 3 & 13 & 1 & 17 \\
& \emph{fr} & 17 & 191 & 7 & 215 \\
& $\overline{\emph{fr}}$ & 5.7 & 14.7 & 7 & 12.9 \\
& $\sigma$ & 0.9 & 24.7 & 0 & 21.9 \\
\hline

\end{tabular}
\end{table}

The Syntest-JavaScript dataset contains the most crashes of the datasets within CrashJS.
Most of these crashes are in the Express (86) and Moment (86) projects, with the least in the Lodash (11) project.
The complexity of the crashes within the Syntest-JavaScript dataset are similar between the projects; 
JavaScript-Algorithms is the most complex with an average of 5.6 frames per stack, while Commander is the least complex with an average of 4.0.
The overall dataset has an average of 5.1 frames per stack with a standard deviation of 3.4.

The majority of the crashes within the Syntest-JavaScript dataset are TypeErrors (257).
If we consider that Syntest-JavaScript crashes are generated from the test generation tool using an evolutionary approach for a dynamically typed language, it becomes clear why the majority are TypeErrors.
Syntest-JavaScript cannot rely on static types to ensure the correct type of arguments are used for function calls.
This means that often incorrect types are used, resulting in the TypeErrors represented in the Syntest-JavaScript dataset.

\begin{table}[htb]
\centering
\caption{Syntest-JavaScript Crash Statistics}
\label{T:syntestStats}
\begin{tabular}{ | c  c | c | c | c | } 
 \hline
 Project &  & TE & Other & Total \\
 \hline
 \hline
commander & \emph{st} & 36 & 2 & 38 \\
& \emph{fr} & 145 & 6 & 151 \\
& $\overline{\emph{fr}}$ & 4.0 & 3.0 & 4.0 \\
& $\sigma$ & 1.5 & 1.0 & 1.5 \\
\hline
express & \emph{st} & 84 & 2 & 86 \\
& \emph{fr} & 421 & 11 & 432 \\
& $\overline{\emph{fr}}$ & 5.0 & 5.5 & 5.0 \\
& $\sigma$ & 2.0 & 0.5 & 1.9 \\
\hline
javascript-algorithms & \emph{st} & 40 & 14 & 54 \\
& \emph{fr} & 197 & 103 & 300 \\
& $\overline{\emph{fr}}$ & 4.9 & 7.4 & 5.6 \\
& $\sigma$ & 2.1 & 11.9 & 6.4 \\
\hline
lodash & \emph{st} & 11 & 0 & 11 \\
& \emph{fr} & 48 & 0 & 48 \\
& $\overline{\emph{fr}}$ & 4.4 &  0 & 4.4 \\
& $\sigma$ & 1.2 & 0.0 & 1.2 \\
\hline
moment & \emph{st} & 86 & 0 & 86 \\
& \emph{fr} & 462 & 0 & 462 \\
& $\overline{\emph{fr}}$ & 5.4 & 0 & 5.4 \\
& $\sigma$ & 2.2 & 0.0 & 2.2 \\
\hline
Total & \emph{st} & 257 & 18 & 275 \\
& \emph{fr} & 1273 & 120 & 1393 \\
& $\overline{\emph{fr}}$ & 5.0 & 6.7 & 5.1 \\
& $\sigma$ & 2.1 & 10.6 & 3.4 \\
\hline

\end{tabular}
\end{table}

An interesting comparison between the GitHub crashes and BugsJS crashes arises when we consider the Express dataset for these two sources.
As discussed previously, Express crashes within the BugsJS dataset have an average of 54 frames per crash (the most complex of CrashJS), while Express crashes within the GitHub dataset have an average of 8.8 (the least complex for the GitHub dataset).
This is, in part, due to the BugsJS crashes using headers, while the GitHub crashes do not.
In Express the handling of headers is complex and requires calls to multiple handlers and other functions within the Express router, thus leading to the significantly larger stack traces within the BugsJS dataset.

\begin{table}[htb]
\centering
\caption{Overall Error Statistics}
\label{T:errorTypes}
\begin{tabular}{ | c | c |  } 
 \hline
 Error & Total  \\
 \hline
 \hline
TypeError & 308 \\
Error & 64 \\
AssertionError & 51 \\
RangeError & 12 \\
SyntaxError & 6 \\
Other & 12 \\ 
\hline

\end{tabular}
\end{table}

\subsection{Program Complexity}

The measure of program complexity we have used is the Cyclomatic Complexity Number (CCN) for each program within CrashJS.
A CCN is calculated from a program using the Control Flow Graph and the formula: \(M = E - N + 2P\) where \(E\) is the number of edges, \(N\) is the number of nodes, and \(P\) is the number of connected components within the graph.
To calculate the CCN for each crash we used the \emph{complexity-report}\footnote{https://github.com/escomplex/complexity-report} JavaScript tool.
The collected CCNs give us a measure of the complexity of each target program for each crash within CrashJS, which, when combined with the crash complexity, gives an overall estimate of the difficulty of reproduction for each crash.

To give a quantitative metric for us to compare the complexities of the datasets we use the formula:
\[C_o = (F_p / F_{max}) + (CCN_p / CCN_{max})\] 
Where \(C_o\) is the overall complexity, \(F_p\) is the average number of frames for the project, \(F_{max}\) is the highest average frames for all of CrashJS (54), \(CCN_p\) is the CCN for the project, and \(CCN_{max}\) is the highest CCN for all of CrashJS (5.2).
We have used this formula as it gives a normalised result across CrashJS with neither crash complexity nor project complexity outweighing the other.
This allows us to make direct comparisons between projects and datasets within CrashJS.
Table~\ref{T:crashjsComplexity} shows the \(F_p\), \(C_p\), and calculated \(C_o\) values for all datasets within CrashJS.

\begin{figure}[htb]
    \includegraphics[width=0.8\linewidth]{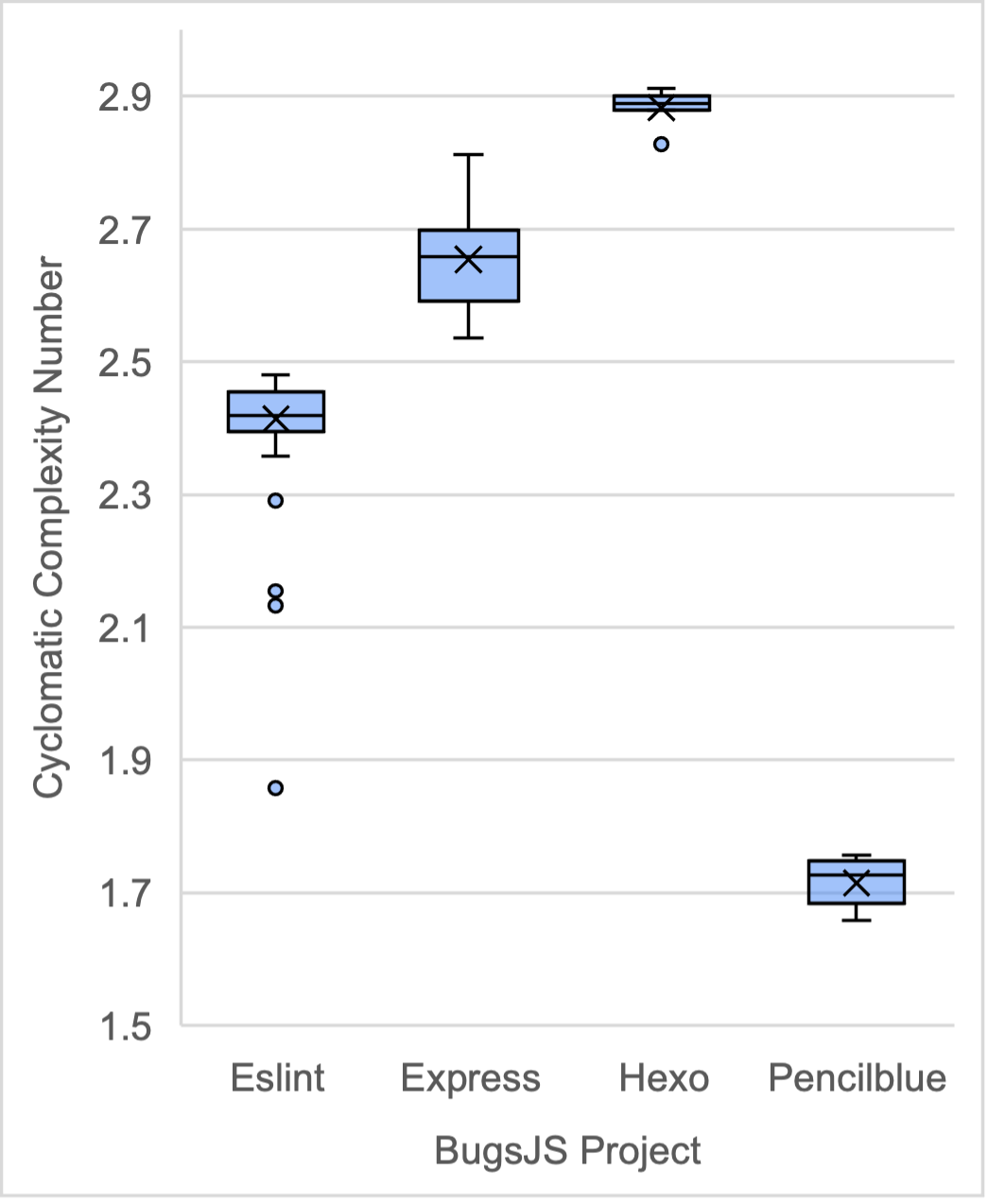}
    \caption{BugsJS CCN Distribution}
    \label{F:bugsjsCCNs}
\end{figure}

Figure~\ref{F:bugsjsCCNs} shows the distribution of CCNs for the BugsJS dataset within CrashJS.
As can be seen, Pencilblue is the least complex project, while Hexo is the most complex.
Eslint has the largest range of complexities, but also significantly more crashes than the other projects, so this is to be expected.
Express contains the next most crashes and has the next widest range of complexities for the project.

As seen in Table~\ref{T:crashjsComplexity}, the calculated \(C_o\) values for the BugsJS projects show Pencilblue, Eslint, and Hexo have relatively low complexity scores compared to Express, although Eslint, Express, and Hexo have similar \(CCN_p\) scores.
In this case, it is clear that the significantly larger Express stack traces skew the complexity of the Express crashes. 
Because of this, BugsJS has the highest overall \(C_o\) of 0.73.

\begin{figure}[htb]
    \includegraphics[width=0.8\linewidth]{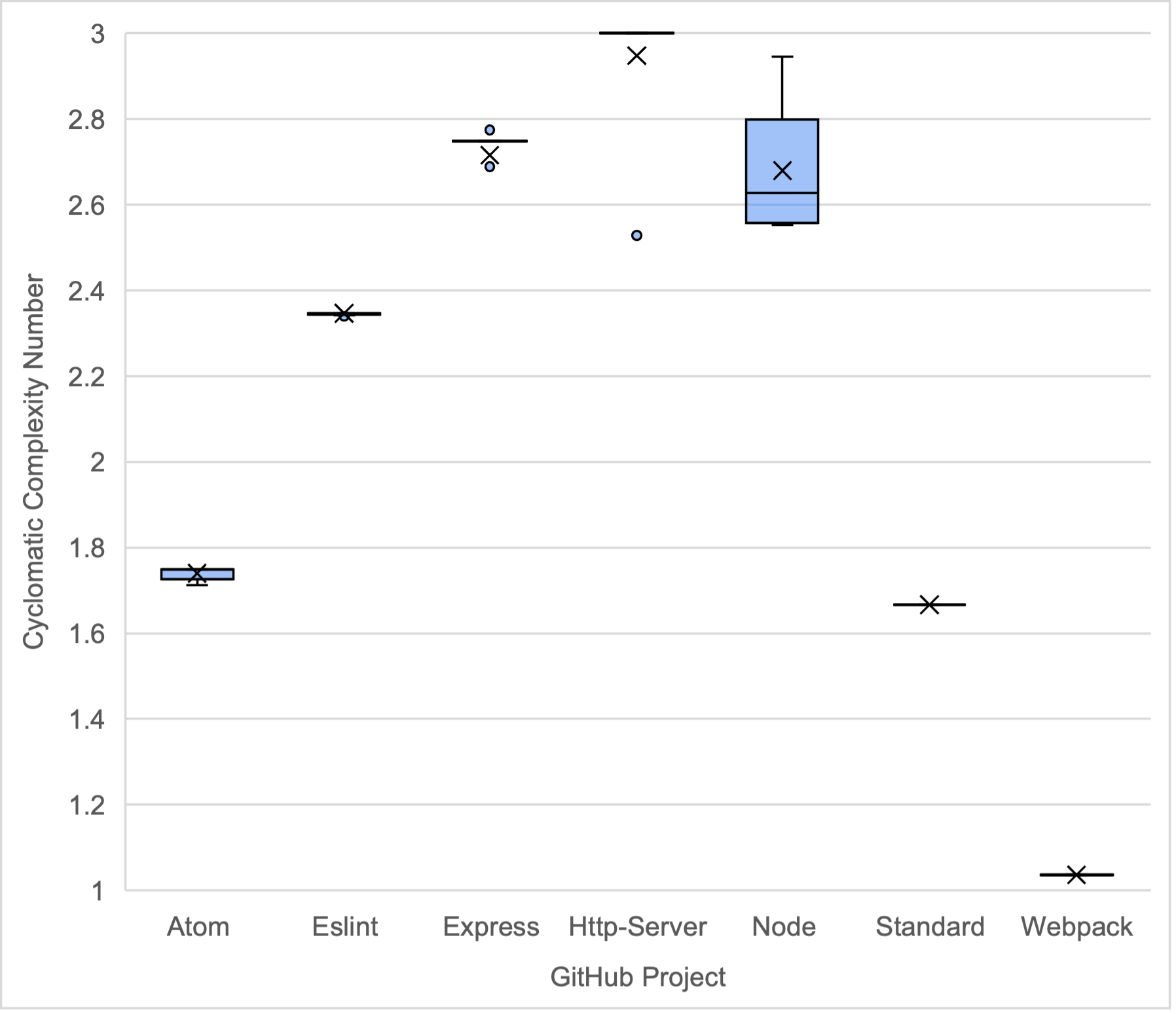}
    \caption{GitHub Crash CCN Distribution}
    \label{F:githubCCNs}
\end{figure}

Figure~\ref{F:githubCCNs} shows a similar spread of CCNs in the GitHub dataset as the BugsJS dataset.
However, the Webpack project has the lowest CCNs of all projects within CrashJS.
This can be seen in Figure~\ref{F:githubCCNs} by the clustering just above 1.0.
Interestingly, there is a significant difference between the spreads of CCN for the Eslint project in the GitHub dataset compared to the BugsJS dataset.
This is likely due to the collection methods for the GitHub dataset.
As the crashes were collected from the GitHub issue tracker for Eslint, the crashes are clustered around similar versions, whereas the BugsJS collection method sampled a much larger range of bugs from within GitHub.

\begin{figure}[htb]
    \includegraphics[width=0.8\linewidth]{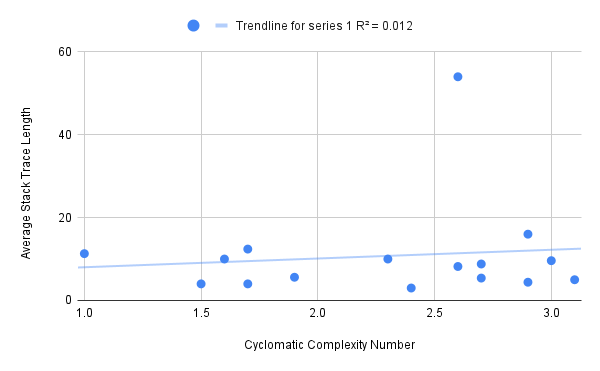}
    \caption{Comparison of Crash and Program Complexity}
    \label{F:complexityComparison}
\end{figure}

As seen in Table~\ref{T:crashjsComplexity}, the majority of projects within the GitHub dataset have similar \(C_o\) values.
Interestingly, the projects with higher \(F_p\) values tend to have lower \(CCN_p\) values and vice versa.
This seems to result in the \(C_o\) values remaining similar.
However, Webpack has a significantly lower \(CCN_p\) than all other projects, resulting in the lowest \(C_o\) value for the GitHub dataset of 0.41.
Further investigation into the possibility of a correlation between crash complexity and program complexity found no correlation between these in our benchmark with an R\textsuperscript{2} value of 0.012 and no discernible shape (Figure~\ref{F:complexityComparison}).

\begin{figure}[htb]
    \includegraphics[width=0.8\linewidth]{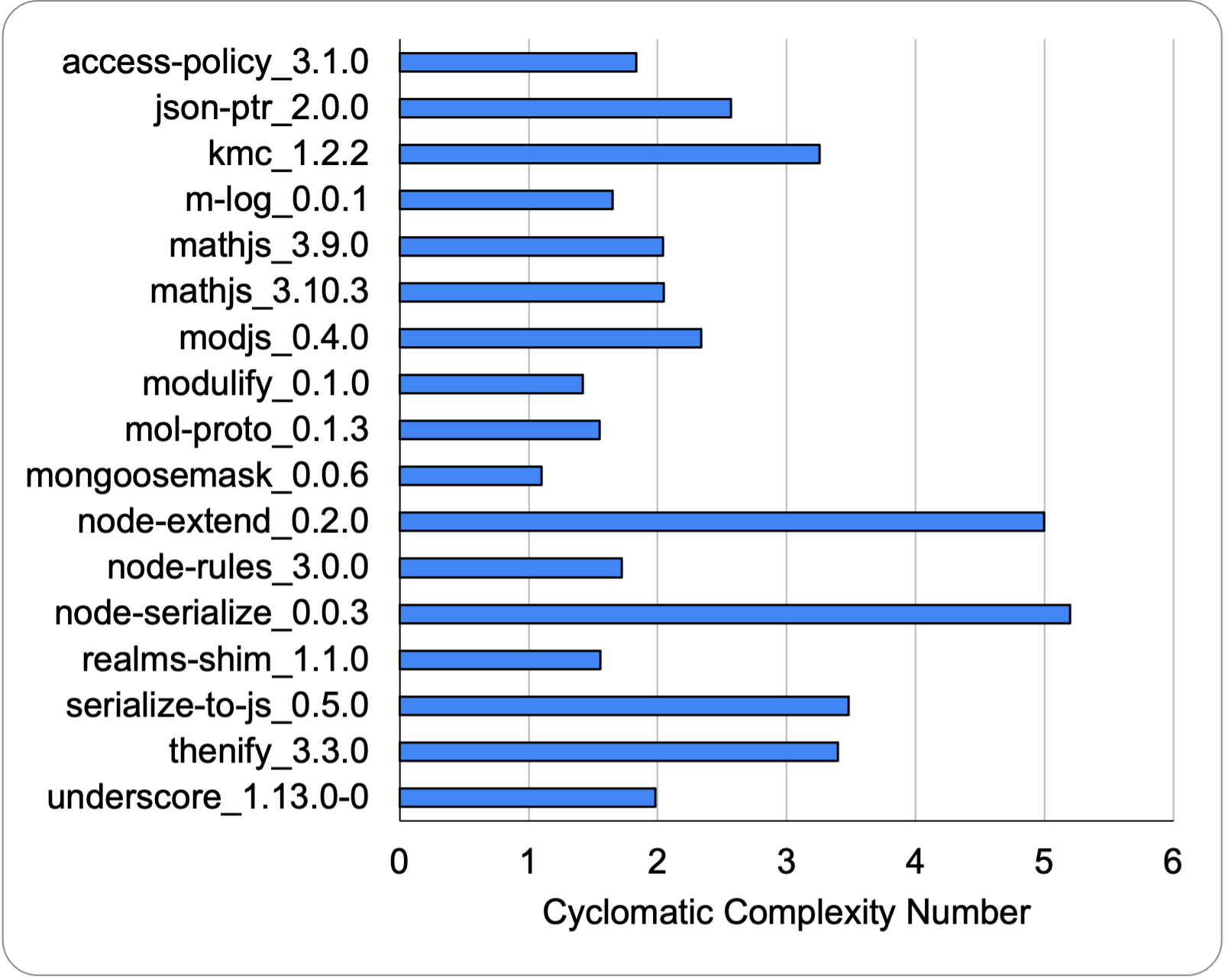}
    \caption{SecBench.js CCNs}
    \label{F:secbenchCCNs}
\end{figure}

As seen in Figure~\ref{F:secbenchCCNs}, there is a similar distribution of CNN values between approximately 1 and 3.
However, there are two significant outliers in node-extend and node-serialize, with values around 5.
These two outliers are the most complex CNN values in CrashJS and could provide insight into ACR tools' capabilities to reproduce crashes in higher-complexity programs.

Table~\ref{T:crashjsComplexity} shows that the SecBench.js dataset is the second-most complex dataset in CrashJS with an overall complexity value of 0.72.
Only BugsJS is more complex, with a value of 0.73.
However, as these values are so similar, the results from ACR tools on these two datasets could prove interesting considering the significant difference in the size of the datasets (90 for BugsJS, 17 for SecBench.js).

\begin{table}[htb]
\centering
\caption{CrashJS Overall Complexity}
\label{T:crashjsComplexity}
\begin{tabular}{ | c c | c | c | c |  } 
 \hline
 Dataset & Project & \(F_p\) & \(CCN_p\) & \(C_o\)  \\
 \hline
 \hline
BugsJS & eslint & 3.0 & 2.41 & 0.52 \\
& express & 54 & 2.65 & 1.51 \\
& hexo & 16 & 2.88 & 0.85 \\
& pencilblue & 4 & 1.71 & 0.40 \\
\hline
& Overall & 14.4 & 2.43 & 0.73 \\ 
\hline
\hline
GitHub & atom & 12.4 & 1.74 & 0.56 \\
& eslint & 10 & 2.35 & 0.64 \\
& express & 8.8 & 2.71 & 0.68 \\
& http-server & 9.6 & 2.95 & 0.75 \\
& node & 8.2 & 2.68 & 0.67 \\
& standard & 10 & 1.67 & 0.51 \\
& webpack & 11.3 & 1.04 & 0.41 \\
\hline
& Overall & 10.2 & 2.19 & 0.61 \\
\hline
\hline
SecBench.js & Overall & 12.9 & 2.48 & 0.72 \\ 
\hline
\hline
Syntest-JS & commander & 4.0 & 2.41 & 0.54 \\
& express & 5.0 & 3.05 & 0.68 \\
& js-algorithms & 5.6 & 1.93 & 0.47 \\
& lodash & 4.4 & 2.91 & 0.64 \\ 
& moment & 5.4 & 2.73 & 0.63 \\
\hline
& Overall & 5.1 & 2.61 & 0.60 \\
\hline
\hline
CrashJS & Overall & 8.01 & 2.50 & 0.63 \\
\hline

\end{tabular}
\end{table}

As previously discussed, all crashes for each project within the Syntest-JavaScript dataset are the same version.
This makes the analysis of CCNs for the dataset significantly simpler.
Interestingly, the CCN for the Express project shows that the version used for Syntest-JavaScript is one of the more complex versions as compared to the BugsJS and GitHub datasets.
Overall, this dataset has similar CCNs to the other datasets, with a slightly higher range beginning just under 2 and finishing just above 3.

\begin{figure}[htb]
    \includegraphics[width=0.8\linewidth]{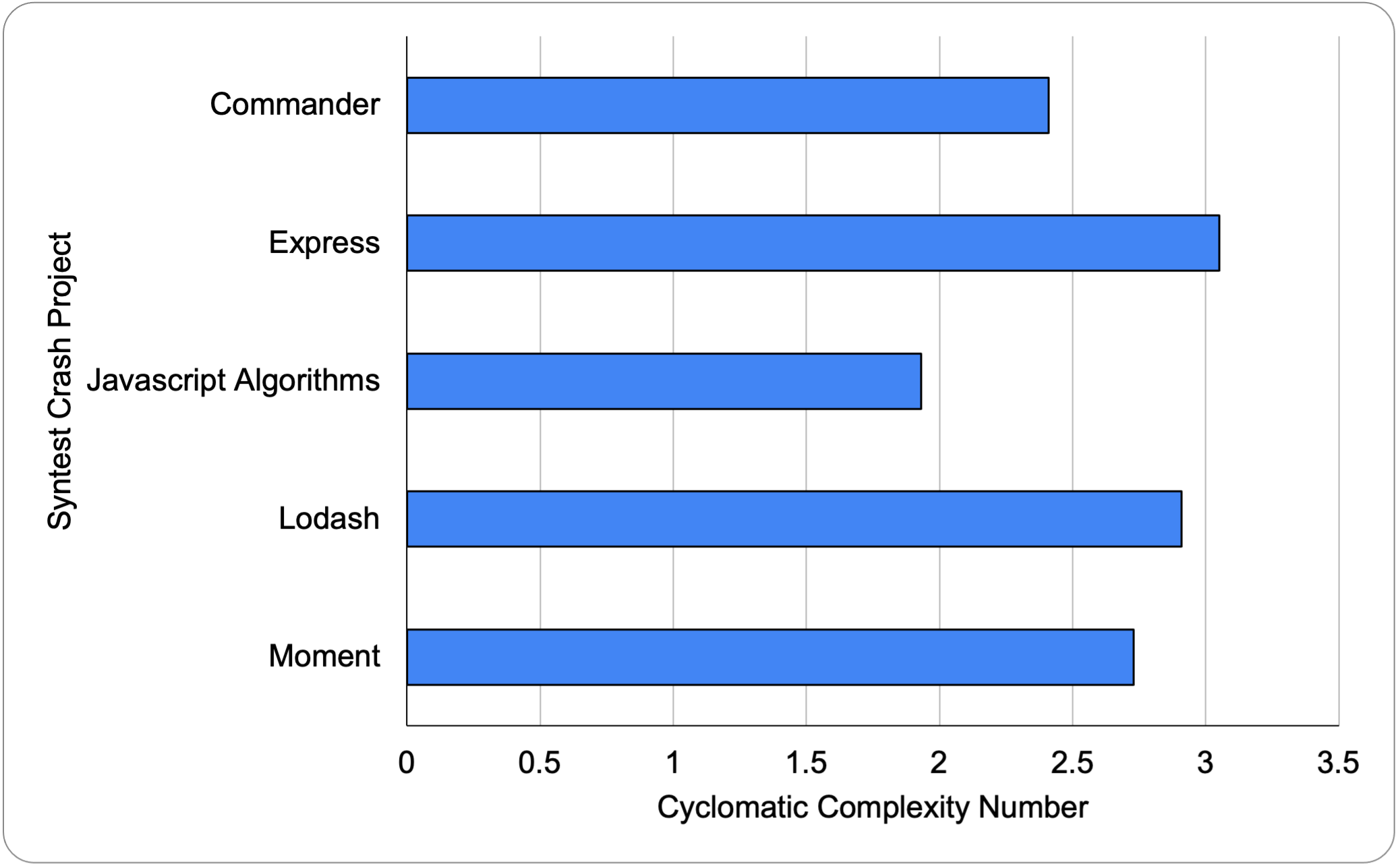}
    \caption{Syntest-JavaScript CCNs}
    \label{F:syntestCCNs}
\end{figure}

Table~\ref{T:crashjsComplexity} shows the Syntest-JavaScript dataset has the lowest overall complexity in CrashJS with an overall \(C_o\) value of 0.60.
This value is only just lower than the GitHub dataset (0.61), however, the Syntest-JavaScript dataset is significantly larger (275 crashes vs 71).
As previously discussed, the majority of errors in this dataset are TypeErrors occurring from randomly generated input data.
We believe this will also lower the complexity of this dataset, providing crashes which early-stage JavaScript ACR tools can use to understand what aspects of their approaches work and how to optimise for more complex crashes.

\section{Conclusions}

Currently, there are no benchmarks for Automated Crash Reproduction tools targeting JavaScript programs.
Several ACR tools have previously been created and tested for C, Java, and Python with varying levels of benchmarks.
Other benchmarks for JavaScript include BugsJS, a collection of bugs from JavaScript programs, SecBench.js, a collection of security vulnerabilities, and the Syntest-JavaScript benchmark, a collection of JavaScript programs to target for automated test generation using the Syntest-JavaScript tool.
We propose CrashJS, the first benchmark specifically for ACR tools targeting JavaScript programs.
CrashJS comprises 453 crashes from four main sources: BugsJS, GitHub, SecBench.js, and Syntest-JavaScript.
The GitHub crashes have been collected from the issue trackers of popular Node.js projects on GitHub.
The other 3 datasets all leverage aspects of the BugsJS, SecBench.js, and Syntest-JavaScript benchmarks to extract crashes for CrashJS.
The crashes collected have a variety of complexity, both in the length of stack trace to be reproduced and the cyclomatic complexity of the target project for the crash.
We believe that CrashJS will allow for fair comparisons between forthcoming JavaScript ACR tools and provide insight into what aspects of ACR approaches provide the most successful reproductions of JavaScript crashes.

\newpage
\bibliographystyle{ACM-Reference-Format}
\bibliography{references}


\end{document}